\documentclass[aps,preprint,showpacs,amsmath,superscriptaddress]{revtex4-1}

\usepackage{hyperref}
\usepackage{siunitx}
\usepackage{graphicx}

\newcommand{\tm}{\mathrm{TM}}
\newcommand{\te}{\mathrm{TE}}
\newcommand{\tr}{\mathrm{tr}}
\newcommand{\w}{\omega}
\newcommand{\ti}{\theta_{i}}
\newcommand{\kb}{k_{B}}
\newcommand{\nf}{\widetilde v_{\textrm{F}}}
\newcommand{\siotwo}{Si$\textrm{O}_2$}

\begin{document}


\title{Theory of reflectivity properties of graphene-coated material plates}


\author{G.\;L. Klimchitskaya}
\affiliation{
    Central Astronomical Observatory at Pulkovo of the Russian Academy of
    Sciences,
    Saint Petersburg,
    196140,
    Russia
}
\affiliation{
    Institute of Physics, Nanotechnology and Telecommunications,
    Peter the Great Saint Petersburg Polytechnic University,
    Saint Petersburg,
    195251,
    Russia
}

\author{C.\;C. Korikov}
\affiliation{
    Institute of Physics, Nanotechnology and Telecommunications,
    Peter the Great Saint Petersburg Polytechnic University,
    Saint Petersburg,
    195251,
    Russia
}

\author{V.\;M. Petrov}
\affiliation{
    Institute of Physics, Nanotechnology and Telecommunications,
    Peter the Great Saint Petersburg Polytechnic University,
    Saint Petersburg,
    195251,
    Russia
}


\begin{abstract}
The theoretical description for the reflectivity properties of dielectric, metal and semiconductor plates coated with
graphene is developed in the framework of the Dirac model. Graphene is described by the polarization tensor allowing
the analytic continuation to the real frequency axis. The plate materials are described by the frequency-dependent
dielectric permittivities. The general formulas for the reflection coefficients and reflectivities of the graphene-coated plates, as well
as their asymptotic expressions at high and low frequencies, are derived. The developed theory is applied to the graphene-coated
dielectric (fused silica), metal (Au and Ni), and semiconductor (Si with various charge carrier concentrations) plates. In all these cases
the impact of graphene coating on the plate reflectivity properties is calculated over the wide frequency ranges. The obtained results can be used
in many applications exploiting the graphene coatings, such as the optical detectors, transparent conductors, anti-reflection surfaces etc.
\end{abstract}


\maketitle

\section{\label{sec:intro}Introduction}

During the last few years, graphene has attracted a lot of experimental and theoretical attention in both
fundamental physics and technological applications~\cite{c01}. Being a two-dimensional sheet of carbon atoms,
graphene possesses unique electrical, mechanical and optical properties. At energies below a few eV
these properties are well described by the Dirac model, which assumes the linear dispersion relation for
massless quasiparticles moving with a  Fermi velocity rather than with the speed of light~\cite{c01,c02}.
New important physics was already revealed in the interaction of graphene with strong magnetic field~\cite{c03}, with the
field of an electrostatic potential barrier~\cite{c04}, and with space homogeneous constant and time-dependent electric fields~\cite{c05,c06,c07}.
Considerable attention has been given to the study of van der Waals and Casimir forces between two graphene sheets and
between a graphene sheet and a plate made of some ordinary material~\cite{c08,c09,c10,c11,c12,c13}. These investigations are closely related to
the subject of the present paper because the Lifshitz theory expresses the force value via the reflection coefficients of interacting
surfaces~\cite{c13a} (through calculated along the imaginary rather than the real frequency axis).

The most fundamental formalism for calculation of the van der Waals and Casimir interaction in layer systems including graphene exploits the
polarization tensor in $(2+1)$-dimensional space-time~\cite{c14,c15,c16,c17,c18}. This formalism was shown~\cite{c19} to be equivalent to
the formalisms which use the spatially nonlocal electric susceptibilities (polarizabilities) of graphene and the density-density
correlation functions in the random-phase approximation. The latter are directly connected with the in-plane and out-of-plane
conductivities of graphene. The Lifshitz theory with reflection coefficients expressed via the polarization tensor was compared~\cite{c20,c21}
with the measurement data for the gradient of the Casimir force between an Au-coated sphere and a graphene-coated plate~\cite{c22},
and demonstrated a very good agreement. It should be mentioned that the previously used hydrodynamic model of graphene was shown to be excluded by
the data~\cite{c23}.

The reflectivity properties of graphene at real frequencies were investigated using the local model for the in-plane (longitudinal) graphene
conductivity~\cite{c24,c25}. As a result, the reflectivity of transverse magnetic (TM), i.e., {\it p}-polarized, electromagnetic waves on graphene
was found at both low~\cite{c24} and high~\cite{c25} frequencies. Reference~\cite{c26} attempted to apply the polarization
tensor of Refs.~\cite{c14,c15} to calculate the reflectivities for both the TM and TE (i.e., transverse electric or {\it s}-polarized)
electromagnetic waves on graphene in
the framework of the Dirac model at any nonzero temperature. However, only the partial results were obtained limited to the case of sufficiently high
frequencies, where one can omit the temperature-dependent part of the polarization tensor. The reason is that the latter part found in
Ref.~\cite{c15} turned out to be correct only at the pure imaginary Matsubara frequencies
and it does not allow analytic continuation to the whole plane of complex frequencies including the real frequency axis.

The complete theory for the reflectivity of graphene over the entire real frequency axis was developed in Ref.~\cite{c27}. For this
purpose another representation for the polarization tensor of graphene was derived, which takes the same values as that
of Ref.~\cite{c15} at the imaginary Matsubara frequencies, but can be analytically continued to the whole frequency
axis complying with all physical requirements. Using this representation, the analytic asymptotic expressions for the reflection coefficients and
reflectivities of graphene at low and high frequencies were obtained for both TM and TE polarizations of the electromagnetic field~\cite{c27}.
The new representation was also used  to investigate the origin of large thermal effect arising in the Casimir interaction between two graphene
sheets~\cite{c28}.

In the present paper, we develop the complete theory allowing calculation of the reflection coefficients and reflectivities
of graphene-coated material plates. This theory exploits the polarization tensor of graphene derived in Ref.~\cite{c27} and, thus, is based on
first principles of quantum electrodynamics. In addition to great fundamental interest, such kind theory is much needed in numerous
technological applications of graphene. Graphene coatings are already used to increase the efficiency of light absorption on optical metal
surfaces~\cite{c29}. This is important for the optical detectors. Graphene-coated substrates have potential applications as transparent
electrodes (see Ref.~\cite{c30} where graphene-silica thin films are used as transparent conductors). Deposition of graphene on silicon
substrates provides an excellent anti-reflection, which is important for solar-cells~\cite{c31}. One could also mention
that graphene coating on metal surfaces is employed for corrosion protection~\cite{c32}. In all these cases, the developed theory can be used to
calculate the effect of graphene coating.

The paper is organized as follows. In Sec.~\ref{sec:analytic} we derive the exact formulas for the reflection coefficients and reflectivities
of the material plate coated with graphene and their asymptotic expressions in the cases of low and high frequencies. In Sec.~\ref{sec:ref_gs}
the developed formalism is applied to the case of graphene-coated amorphous \siotwo\;(silica) plate. We show that at high frequencies the graphene
coating slightly increases the reflectivity of silica (which is small in any case). It leads, however, to some increase of the Brewster angle.
At low (microwave) frequencies graphene coating significantly increases the reflectivities of a silica plate. Section~\ref{sec:ref_gm} is
devoted to the study of reflectivities of graphene-coated metals (Au and Ni).
According to our results, for each metal there exists a frequency range where the graphene coating leads to some decrease of the reflectivities.
In Sec.~\ref{sec:ref_s} the reflectivities of graphene-coated Si with different concentrations of charge carriers are considered. It is shown that
at  high and moderate concentrations the graphene coating does not influence the reflectivities of Si at high frequencies.
At low frequencies and moderate or small concentrations it is always possible to find the range of frequencies, where the presence of graphene coating increases the reflectivities. Note also that the graphene coating
increases the value of the incidence angle such that  the polarization of incident light reaches its maximum value. This increase is larger for
Si plates with smaller concentrations of charge carriers. Finally, Sec.~\ref{sec:ref_cd} contains our conclusions and discussion.

\section{\label{sec:analytic}Analytic expressions for the reflection coefficients of graphene-coated plates}

In this section, we present general theoretical results for the reflectivity properties of thick material plate (semispace)
coated with the layer of graphene. We assume that the material of the plate is nonmagnetic and is described by the frequency-dependent dielectric
permittivity $\epsilon(\omega)$. Graphene is considered as a pristine (gapless) one (the generalization for the case of a
nonzero gap is straightforward). It is described by the polarization tensor $\Pi_{ik} (\omega,k_\perp)$ with $i,k=0,1,2$ derived in
Ref.~\cite{c27}, where $k_\perp$ is the magnitude of the wave vector projection on the plane of graphene.

The reflection coefficients on a graphene-coated plate, where graphene is described by the polarization tensor and plate material by
the dielectric permittivity, were obtained in Ref.~\cite{c20} at the imaginary Matsubara frequencies.
Taking into account that the polarization tensor of Ref.~\cite{c27} is valid also at the real frequencies, one can leave the same derivation
unchanged. We only take into account that at real frequencies the mass-shell equation for photons is satisfied resulting in the equation
\begin{equation}
k_\perp=\frac{\omega}{c}\sin{\ti},
\label{eq:01}
\end{equation}
where $\ti$ is the angle of incidence. This allows to express the reflection coefficients in Eqs.~(9) and~(19) of Ref.~\cite{c20} and the
polarization tensor in terms of $\ti$ rather than $k_\perp$. As a result, the reflection coefficients of graphene-coated plates for TM and TE
polarizations of graphene-coated plates for TM and TE polarizations of the electromagnetic field take the form
\begin{align}
\begin{split}
&R^{(g,p)}_{\textrm{TM}}(\omega,\ti)
=
\frac{
    \epsilon(\omega) \cos{\ti} - \sqrt{\epsilon(\omega)-\sin^2{\ti}}
    \left[
        1- \Pi_{00}^{*} (\omega,\ti)
    \right]
}
{
    \epsilon(\omega) \cos{\ti} + \sqrt{\epsilon(\omega)-\sin^2{\ti}}
    \left[
        1+ \Pi_{00}^{*} (\omega,\ti)
    \right]
},\\
&R^{(g,p)}_{\textrm{TE}}(\omega,\ti)
=
\frac{
    \cos{\ti} - \sqrt{\epsilon(\omega)-\sin^2{\ti}}
    - \Pi^{*} (\omega,\ti)
}
{
    \cos{\ti} + \sqrt{\epsilon(\omega)-\sin^2{\ti}}
    + \Pi^{*} (\omega,\ti)
}.
\end{split}
\label{eq:02}
\end{align}
Here  the following notations are introduced:
\begin{align}
\begin{split}
&\Pi_{00}^{*} (\omega,\ti) = - \frac{i c}{\hbar \omega} \frac{\cos\ti}{\sin^2\ti} \Pi_{00} (\omega,\ti),\\
&\Pi^{*} (\omega,\ti) = \frac{i c^3}{\hbar \omega^3} \frac{1}{\sin^2\ti} \Pi (\omega,\ti),
\end{split}
\label{eq:03}
\end{align}
where
\begin{equation}
\Pi(\w,0) = \frac{\w^2}{c^2}
\left[
\sin^2\ti \Pi_{\tr} (\w,\ti) + (1-\sin^2\ti) \Pi_{00} (\w,\ti)
\right]
\label{eq:04}
\end{equation}
and $\Pi_{\tr} (\w,\ti) = \Pi_{k}^{k} (\w,\ti)$ is the trace of the polarization tensor.

It is convenient to represent the quantities $\Pi_{00} (\w,\ti)$ and $\Pi_{tr} (\w,\ti)$ in the form
\begin{align}
\begin{split}
&\Pi_{00} (\w,\ti) = \Pi_{00}^{(0)} (\w,\ti) + \Delta_{T} \Pi_{00} (\w,\ti)\\
&\Pi (\w,\ti) = \Pi^{(0)} (\w,\ti) + \Delta_{T} \Pi (\w,\ti),
\end{split}
\label{eq:05}
\end{align}
where $\Pi_{00}^{(0)}$, $\Pi^{(0)}$ are  the respective quantities at zero temperature and $\Delta_{T} \Pi_{00}$, $\Delta_{T} \Pi$ are
the thermal corrections to them. The polarization tensor of graphene at $T=0$ was found in Ref.~\cite{c14}. Along the frequency axis it takes
the form
\begin{align}
\begin{split}
&\Pi_{00}^{(0)} (\w,\ti) = i \pi \alpha \hbar \frac{\w}{c} \frac{\sin^2 \ti}{u(\ti)},\\
&\Pi^{(0)} (\w,\ti) = -i \pi \alpha \hbar \frac{\w^3}{c^3} \sin^2\ti u(\ti),
\end{split}
\label{eq:06}
\end{align}
where
\begin{equation}
u(\ti) = \sqrt{1-\nf^2\sin^2\ti}
\label{eq:07}
\end{equation}
and $\nf=v_{\textrm{F}}/c$ is the dimensionless Fermi velocity. Note that for graphene $\nf \approx 1/300$, so that the quantity $u$
is approximately equal to unity for all $\ti$.

The explicit expressions for the temperature corrections in Eq.~\eqref{eq:05} valid along the real frequency axis are obtained in
Ref.~\cite{c27} [see Eqs.~(45) and (60) in that paper]. Here it is convenient to rewrite them as the functions of $\w$ and $\ti$ in the form of
integrals over the dimensionless variable $y$:
\begin{align}
\begin{split}
&\Delta_{T}\Pi_{00} (\w,\ti) =
\frac{8 \alpha \hbar \w}{\nf^2 c}
\int_0^\infty
\frac{dy}{e^{\beta y}+1}
\left[
1+ \sum_{\lambda=\pm1}
\frac{X_{\lambda}(\ti,y)}{2u(\ti)}
\right],\\
&\Delta_{T}\Pi (\w,\ti) =
\frac{8 \alpha \hbar \w^3}{\nf^2 c^3}
\int_0^\infty
\frac{dy}{e^{\beta y}+1}
\left\{
    1+
    \frac{u(\ti)}{2}
    \sum_{\lambda=\pm1}
    \left[
        X_{\lambda} (\ti,y)
        +
        \tilde{v}_F^2
        \frac{\sin^2\ti}{X_{\lambda}(\ti,y)}
    \right]
\right\},
\end{split}
\label{eq:08}
\end{align}
where $\beta \equiv \w/(2 \w_{T})$, the thermal frequency is defined as $\w_{T} = \kb T / \hbar$, and $\kb$ is the Boltzmann constant. The quantity
$X_{\lambda}$ is defined diffrently at $\lambda = 1$ and $\lambda=-1$. Thus, at $\lambda=1$ we have~\cite{c27}
\begin{equation}
X_1(\ti,y) = - \left[ u^2(\ti) + 2 y + y^2 \right]^{1/2}
\label{eq:09}
\end{equation}
and at $\lambda=-1$ the result is more complicated~\cite{c27}
\begin{equation}
X_{-1}(\ti,y) =
\begin{cases}
    -\left[u^2(\ti)-2y+y^2\right]^{1/2}, & y \le 1-\nf \sin\ti,\\
    -i\left[-u^2(\ti)+2y-y^2\right]^{1/2}, & 1-\nf \sin\ti < y < 1+\nf \sin\ti,\\
    \left[u^2(\ti)-2y+y^2\right]^{1/2}, & y \ge 1+\nf \sin\ti.
\end{cases}
\label{eq:10}
\end{equation}

We now consider the asymptotic expressions for the polarization tensor and reflectivities of graphene at high and low frequencies with respect to the
thermal frequency $\w_{T}$. At room temperature $T=300$ K the thermal frequency $\w_{T}=3.9\times 10^{13}$ rad/s $\approx$ 0.026 eV.
The high-frequency asymptotic expression is applicable under the condition
\begin{equation}
\w \gg \w_{T}.
\label{eq:11}
\end{equation}
This condition is satisfied for all frequencies $\w>0.26$ eV leading to $\beta \gg 1$ and to the smallness of the temperature
corrections~\eqref{eq:08} relative to the zero-temperature contribution~\eqref{eq:06}.
Taking into account also that $\nf \ll 1$, one obtains
\begin{align}
\begin{split}
&\Pi_{00} (\w,\ti) \approx \Pi_{00}^{(0)} (\w,\ti) \approx i \pi \alpha \hbar \frac{\w}{c} \sin^2\ti,\\
&\Pi (\w,\ti) \approx \Pi^{(0)} (\w,\ti) \approx - i \pi \alpha \hbar \frac{\w^3}{c^3} \sin^2\ti.
\end{split}
\label{eq:13}
\end{align}
Then, the quantities~\eqref{eq:03} entering the reflection coefficients~\eqref{eq:02} take the form
\begin{align}
\begin{split}
&\Pi_{00}^{*} (\w,\ti) \approx \pi \alpha \cos\ti,\\
&\Pi^{*} (\w,\ti) \approx \pi \alpha.
\end{split}
\label{eq:13}
\end{align}

Note that the first of these quantities does not depend on $\w$ and the second one does not depend also on $\ti$. Because of this,
at high frequencies the frequency dependence of the reflection coefficients~\eqref{eq:02}
is completely determined by the dielectric permittivity of a plate.

At the normal incidence ($\ti=0$) from Eqs.~\eqref{eq:02} and~\eqref{eq:13} one obtains
\begin{equation}
R^{(g,p)}_{\tm}(\omega,0) = - R^{(g,p)}_{\te}(\omega,0) =
\frac{n_1(\omega)-1+\pi\alpha+in_2(\w)}
{n_1(\omega)+1+\pi\alpha+in_2(\w)}.
\label{eq:14}
\end{equation}
where $n_1(\w)$ and $n_2(\w)$ are the real and imaginary parts, respectively, of the complex index of refraction of plate material
\begin{equation}
n(\w) = n_1(\w) + i n_2(\w).
\label{eq:15}
\end{equation}
From Eq.~\eqref{eq:14} for the reflectivity of graphene-coated plate in the region of high frequencies at the normal incidence we have
\begin{equation}
\mathcal{R}^{(g,p)}(\omega,0)=\left|R^{(g,p)}_{\tm}(\omega,0)\right|^2 = \left|R^{(g,p)}_{\te}(\omega,0)\right|^2 \approx
\frac{[n_1(\omega)-1+\pi \alpha]^2+n_2^2(\w)}
{[n_1(\omega)+1+\pi \alpha]^2+n_2^2(\w)}.
\label{eq:16}
\end{equation}
This should be compared with the reflectivity of an uncoated plate at the normal incidence
\begin{equation}
\mathcal{R}^{(p)}(\omega,0) =
\frac{[n_1(\omega)-1]^2+n_2^2(\w)}
{[n_1(\omega)+1]^2+n_2^2(\w)}.
\label{eq:17}
\end{equation}

Let us consider now the asymptotic expressions for the polarization tensor and reflectivity of  graphene
at low frequencies satisfying the condition
\begin{equation}
\w \ll \w_{T}.
\label{eq:18}
\end{equation}
In this case, the polarization tensor can be obtained from Eqs. (89) and (95) of Ref.~\cite{c27} taking into account Eq.~\eqref{eq:01} and
the smallness of dimensionless Fermi velocity. The results are
\begin{align}
\begin{split}
&\Pi_{00} (\w,\ti) \approx  - 8 \alpha \ln{2} \frac{\kb T}{c} \sin^2\ti,\\
&\Pi (\w,\ti) \approx   8 \alpha \ln{2} \frac{\kb T \w^2}{c^3} \sin^2\ti.
\end{split}
\label{eq:19}
\end{align}

Substituting Eq.~\eqref{eq:09} in Eq.~\eqref{eq:03}, we find
\begin{align}
\begin{split}
&\Pi_{00}^{*} (\w,\ti) \approx   8 i \alpha \ln{2} \frac{\w_{T}}{\w} \cos\ti,\\
&\Pi^{*} (\w,\ti) \approx  8 i \alpha \ln{2} \frac{\w_{T}}{\w}.
\end{split}
\label{eq:20}
\end{align}

Then, from Eq.~\eqref{eq:02} one obtains the reflection coefficients for a graphene-coated plate at low frequencies.
At the normal incidence from Eqs.~\eqref{eq:02} and~\eqref{eq:20} one arrives at
\begin{equation}
R^{(g,p)}_{\tm}(\omega,0) = - R^{(g,p)}_{\te}(\omega,0) \approx
\frac{n_1(\omega)-1+i\left[n_2(\omega)+8\alpha \frac{\omega_\tau}{\omega} \ln{2} \right]}
{n_1(\omega)+1+i\left[n_2(\omega)+8\alpha \frac{\omega_\tau}{\omega} \ln{2} \right]}.
\label{eq:21}
\end{equation}
Then, the reflectivity at the normal incidence is given by
\begin{equation}
\mathcal{R}^{(g,p)}(\omega,0) \approx
\frac{[n_1(\omega)-1]^2+\left[n_2(\omega)+8\alpha \frac{\omega_\tau}{\omega} \ln{2} \right]^2}
{[n_1(\omega)+1]^2+\left[n_2(\omega)+8\alpha \frac{\omega_\tau}{\omega} \ln{2} \right]^2}.
\label{eq:22}
\end{equation}

Note that in contrast to Eq.~\eqref{eq:16} at low frequencies the reflectivity depends on the temperature. In the absence of graphene
coating one returns back to Eq.~\eqref{eq:17}.

\section{\label{sec:ref_gs}Reflectivity of graphene-coated silica}

Now we apply the developed formalism to the case of amorphous \siotwo\;(silica) plate coated with graphene film.
In all computations we use the optical data for the complex index of refraction of silica~\cite{c33} tabulated in the frequency region 0.00248 eV
$\le \omega \le$ 2000 eV. At smaller frequencies $\omega <$ 0.00248 eV it holds $n_1 = 1.956=$const, $n_2=0$.
We start from the case of high frequencies~\eqref{eq:11}. According to the results of Sec.~\ref{sec:analytic}, this case includes the experimentally
interesting regions of optical and ultraviolet frequencies. The reflectivity of the graphene-coated silica at the
normal incidence as a function of frequency is computed by Eq.~\eqref{eq:16}. The  computational results are presented in Fig.~1 by the line 2.
In the same figure line 1 shows the reflectivity of an uncoated silica plate at the normal incidence computed by Eq.~\eqref{eq:17}.
As can be seen in Fig.~1, in the regions of optical and ultraviolet frequencies the graphene coating slightly (from 6\% to 9\%)
increases the reflectivity of a silica plate at the normal incidence.

Next, we consider the dependence of reflectivities on the angle of incidence at high frequencies. For this purpose, we calculate the
quantities

\begin{equation}
\mathcal{R}^{(g,p)}_{\textrm{TM}(\textrm{TE})}(\omega,\ti)
=
\left|R^{(g,p)}_{\textrm{TM}(\textrm{TE})}(\omega,\ti)\right|^2,
\label{eq:23} 
\end{equation}
where $R^{(g,p)}_{\textrm{TM}(\textrm{TE})}(\omega,\ti)$ are given by Eqs. \eqref{eq:02} and \eqref{eq:13}. The computational results
at $\omega=2$~eV$=3\times10^{15}$ rad/s (visible light) are presented in Fig.~2(a) as a function of $\ti$ by the lower and upper
solid lines (for the TM and TE polarizations, respectively). In the same figure, the dashed lines show the computational results in the
absence of graphene coating. As is seen in Fig.~2(a), the Brewster angle, at which the reflected light is fully (TE) polarized, is slightly different
for the uncoated and graphene-coated silica plates. To make this effect more quantitative, in Fig.~2(b) we plot the reflectivities
$\mathcal{R}_{\textrm{TM}}$
on a larger scale in the vicinity of the Brewster angle (the dashed and solid lines are for an uncoated and for a graphene-coated silica plates,
respectively). As is seen in Fig.~2(b), graphene coating leads to the increase of the Brewster angle from $\theta_{\textrm{B}}=55.5^{\circ}$ to
$\theta_{\textrm{B}}=56.2^{\circ}$.

We are coming now to the asymptotic region of low frequencies~\eqref{eq:18}. In this case the reflectivity of the graphene-coated silica plate at
the normal incidence is computed by Eq.~\eqref{eq:22}. The computational results as a function of frequency are shown in Fig.~3 by the
lines 2 and 3 at $T=75$ K and $T=300$ K, respectively. The line 1 shows the reflectivity of an uncoated silica plate at the normal incidence
computed by Eq.~\eqref{eq:17}. As is seen in Fig.~3, at low frequencies the graphene coating leads to significant (up to an order of magnitude)
increase of the reflectivity of silica at the normal incidence. It is seen also that, unlike the case of high frequencies, there is strong
dependence of the reflectivity on temperature. Thus, at $\omega=0.1, 0.5$, and 1 eV the reflectivity at $T=300$ K is larger than that at $T=75$ K
by the factors of 1.88, 2.91, and 1.76, respectively.

The dependence of the TM and TE reflectivities on the angle of incidence is easily obtainable by using Eqs.~\eqref{eq:02},~\eqref{eq:20} and
\eqref{eq:23}. In the range of low frequencies, however, the full polarization of an incident light is not achieved at any angle.

\section{\label{sec:ref_gm}Reflectivity of graphene-coated metal surfaces}

In this section, we consider the reflectivity properties of metal plates coated with graphene layer. We start with the case of Au plate. The
complex index of refraction of Au is tabulated in Ref.~\cite{c33} in the frequency region 0.125 eV $< \omega <$ 9919 eV. In the most part of
this region (specifically, at $\omega>0.26$ eV) one can use the asymptotic expression at high frequencies to calculate the reflectivities
of the graphene-coated Au plate. The reflectivity of an uncoated Au plate at the normal incidence is shown in Fig.~4(a) as
a function of frequency. Computations were performed by using Eq.~\eqref{eq:17}.

Note that the influence of graphene coating on metal surfaces is smaller than that on dielectric ones. Because of this it is not convenient
to show the reflectivity of the graphene-coated Au plate in the same figure as the uncoated one. Instead, we calculate the relative quantity

\begin{equation}
\delta \mathcal{R}_{\textrm{Au}} (\omega, \ti=0) =
\frac{\mathcal{R}^{(g,p)}_{\textrm{Au}}(\omega,0)-\mathcal{R}^{(p)}_{\textrm{Au}}(\omega,0)}
{\mathcal{R}^{(p)}_{\textrm{Au}}(\omega,0)},
\label{eq:24}
\end{equation}
where $\mathcal{R}^{(g,p)}_{\textrm{Au}}(\omega,0)$ in the high-frequency region $\omega>0.26$ eV was computed using Eqs.~\eqref{eq:16} and~\eqref{eq:23}. At
lower frequencies the exact expressions~\eqref{eq:05},~\eqref{eq:06} and \eqref{eq:08} for the polarization tensor have
been used. At $\omega < 0.125$ eV the optical data were  extrapolated by means of the Drude model~\cite{c34}. The computational results
for the quantity~\eqref{eq:24} versus frequency are presented in Fig.~4(b) in the frequency range from 0.125 eV to 14 eV.

As is seen in Fig.~4(b), for an Au plate the reflectivity at the normal incidence becomes smaller due to the presence of graphene
coating (this is opposite to the case of silica considered in Sec.~\ref{sec:ref_gs}). The single exception from this
observation is the frequency interval 6.7 eV $<\omega<$ 7.5 eV, where $\delta \mathcal{R}_{\textrm{Au}} (\omega, \ti=0) > 0$. The largest in
magnitude influence of graphene coating $|\delta \mathcal{R}_{\textrm{Au}} (\omega, \ti=0)| = 1.4\%$ takes place at $\omega = 2.4$ eV $=3.65\times 10^{15}$
rad/s. It should be stressed also that in the whole region $\omega<0.26$ eV (the moderate and low frequencies)
$\delta \mathcal{R}_{\textrm{Au}} (\omega, \ti=0)=0$ with high accuracy, i.e., the graphene coating does not influence the reflectivity
properties. As a result, the reflectivity of the graphene-coated Au plate does not depend on temperature. These observations are in agreement
with the fact that graphene coatings do not influence the van der Waals and Casimir forces between metal plates~\cite{c21}.

It is instructive also to consider the dependences of the TM and TE reflectivities of a graphene-coated Au plate on the angle of incidence.
The computational results obtained by Eqs.~\eqref{eq:02},~\eqref{eq:13} and~\eqref{eq:23} at $\omega=2.4$ eV $=3.65 \times 10^{15}$ rad/s are
shown in Fig.~5 (the lower and upper solid lines are for the TM and TE polarizations, respectively).
The dashed lines show the computational results for an uncoated Au plate. As is seen in Fig.~5, graphene coating only slightly influences the
angle dependences of the reflectivities. The full polarization of an incident light is not achieved at any frequency.

The results for the reflectivity of graphene-coated magnetic metals are somewhat different from the case of nonmagnetic ones. As an example,
here we consider Ni. The optical data for the complex index of refraction of Ni in the frequency region 0.1 eV $< \omega <$ 9919 eV are also
contained in Ref.~\cite{c33}. The reflectivity of an uncoated Ni plate at the normal incidence is calculated by Eq.~\eqref{eq:14}
and the results are presented
in Fig.~6(a) in the frequency region from $0.1$ eV to $14$ eV.
The comparison with Fig.~4(a) shows that at low frequencies the reflectivity of Ni decreases faster than for Au,
but has no an abrupt fall in the region above 2 eV.

The relative change of reflectivity of the graphene-coated Ni is defined in the same way as that for Au in Eq.~\eqref{eq:24}. It was computed as described above for Au,
and the computational results versus frequency are shown in Fig.~6(b). As can be seen in this figure,
for a Ni plate the reflectivity at the normal incidence becomes smaller due to presence of graphene. This is qualitatively
similar to the case of Au plate. Here, however, the maximum in magnitude influence $|\delta \mathcal{R}_{\textrm{Au}} (\omega, \ti=0)| = 2.5\%$ takes
place not for a visible light (as it holds for Au), but at $\omega=10.4$ eV $=1.58\times 10^{16}$ rad/s, i.e., in the ultraviolet region. Similar to
the  case of Au, at moderate and low  frequencies the graphene coating does not influence the reflectivity of Ni plate (at $\omega < 0.1$ eV the
optical data for Ni were extrapolated using Drude model~\cite{c35}). As a result, it does not depend on the temperature.

\section{\label{sec:ref_s}Reflectivity of graphene-coated S\lowercase{i} with different concentrations of charge carriers}

Now we turn out attention to the graphene-coated Si plates, where Si may have different concentrations of charge carriers due to different dopings.
We start from the case of high-resistivity dielectric Si. In this case, the optical data for the complex index of refraction are again contained
in Ref.~\cite{c33} over the frequency region from 0.004959 eV to 2000 eV. In the region $\omega < 0.004959$ eV one has $n_1=3.416$, $n_2=0$.

In Fig.~7(a) we plot the reflectivity of an uncoated dielectric Si at the normal incidence as a function of frequency
computed by Eq.~\eqref{eq:17} over the entire frequency region. Computations show that in the region of high frequencies defined in Eq.~\eqref{eq:11}
the impact of graphene coating achieves several percent only starting from a few eV.  Because of this, for the characterization of this
impact, it is convenient to use the relative change in the reflectivity of graphene-coated Si plate defined like that for Au
in Eq.~\eqref{eq:24}. In Fig.~7(b) the quantity $\delta \mathcal{R}_{\textrm{Si}} (\ti=0)$ is shown as a function of frequency in the region of
high frequencies from 0.26 eV to 14 eV. Computations were performed by Eq.~\eqref{eq:16}. As is seen in Fig.~7(b), the relative change in the
reflectivity of a graphene-coated Si plate is a monotonously decreasing function of frequency. Its magnitude achieves the
largest value $|\delta \mathcal{R}_{\textrm{Si}} (\omega, \ti=0)| = 6\%$ at $\w=14$~eV. It is notable that the graphene coating
increases the reflectivity of a high resistivity Si plate at optical frequencies, but decreases it at ultraviolet frequencies.

We consider now the region of moderate and low frequencies $\w<0.26$~eV = 260~meV. In this region, the asymptotic case of low frequencies~\eqref{eq:18}
holds for $\w \le 2.6$~meV. Here,  computations can be performed by Eq.~\eqref{eq:22}. In the region from 2.6~meV to 0.26~eV the exact
equations~\eqref{eq:02}--\eqref{eq:08} should be used. The computational results for the reflectivity of graphene-coated
high-resistivity Si plate at the normal
incidence at $T=300$~K is shown in Fig.~7(c) by the line 2 as a function of frequency. In the same figure, the line 1 shows the reflectivity of
an  uncoated high-resistivity Si plate at the normal incidence. The line 1 to within 1\% represents also the reflectivity of the
graphene-coated Si plate at zero temperature. As is seen in Fig.~7(c), the role of graphene coating increases with decreasing frequency.
At $\w=0.01$~meV the reflectivity of the graphene-coated Si is larger than for the uncoated one by the factor of 3.33. Another conclusion is that for
high-resistivity Si the total role of graphene coating is provided by the thermal effect. At $\w>4$~meV both the relative impact of graphene coating
on the reflectivity at the normal incidence and, correspondingly, the thermal effect are in the limits of 1\%. The impact of the graphene
coating again becomes larger in magnitude only in the region of high frequencies [see Fig.~7(b)], where it does not depend on the temperature.

Next, we consider the case of doped low-resistivity Si plates with some concentrations of charge carriers. In this case the dielectric
permittivity of the plate material can be presented in the form

\begin{equation}
\epsilon(\w) = [n_1(\w)+i n_2(\w)]^2 = \epsilon_d(\w)-\frac{\w_p^2}{\w(\w+i\gamma)},
\label{eq:25}
\end{equation}
where $\epsilon_d (\w) = [n_{1d}(\w)+i n_{2d}(\w)]^2$ is the dielectric permittivity of high-resistivity Si and $n_{1d}(\w)$, $n_{2d}(\w)$
are the used above tabulated optical data for its complex index of refraction~\cite{c33}. The free charge carriers are taken into account by means
of the Drude-like term, where $\w_p$ is the plasma frequency and $\gamma$ is the relaxation parameter. As a result, the real and imaginary parts
$n_1(\w)$ and $n_2(\w)$ of the complex refraction index of the low-resistivity Si are calculated from the following system of equations:
\begin{align}
\begin{split}
&n_1^2(\w)-n_2^2(\w) = n_{1d}^2(\w) - n_{2d}^2(\w) - \frac{\w_p^2}{\w^2+\gamma^2},\\
&n_1(\w)n_2(\w) = n_{1d}(\w)n_{2d}(\w) + \frac{\w_p^2\gamma}{2\w(\w^2+\gamma^2)}.
\end{split}
\label{eq:26}
\end{align}

Calculations using Eq.~\eqref{eq:26} show that in the asymptotic region of high frequencies the presence of free charge carriers
leads to only a negligibly small change of the complex index of refraction, as compared to the case of dielectric Si. Because of this,
in the region of high frequencies the influence of graphene coating on the reflectivity of low-resistivity Si plate is again shown by Fig.~7(b).

Now we perform computations of the reflectivity of low-resistivity P-doped Si at the normal incidence in the region of moderate and low frequencies
$\w < 0.26$~eV. As above, for $\w\le2.6$ meV the asymptotic equation~\eqref{eq:18} was used. In the frequency interval 2.6~meV$<$0.26~eV computations
were performed by the exact equations~\eqref{eq:02}--\eqref{eq:08}. The computational results at $T=300$~K as a function of frequency are shown
in Fig.~8 by the solid lines 1, 2, and 3 for the concentrations of charge carriers $N_1 = 5\times10^{14}\text{cm}^{-3}$,
$N_2=5\times10^{16}\text{cm}^{-3}$, and $N_3=5\times10^{17}\text{cm}^{-3}$. At these concentrations, the respective resistivities $\rho_i$ were
obtained from Ref.~\cite{c36} and the Drude parameters were found from the equations~\cite{c37}
\begin{equation}
\w_{p,i} = \frac{e \sqrt{4\pi N_i}}{\sqrt{m^{*}}},\qquad\gamma_i = \frac{\rho_i\w_{p,i}^2}{4\pi},
\label{eq:27}
\end{equation}
where $m^*=0.26 m_e$ is the effective mass, $m_e$ is the electron mass, with the results $\w_{p,1}=2.5\times10^{12}$ rad/s,
$\w_{p,2}=2.5\times10^{13}$ rad/s, $\w_{p,3}=7.8\times10^{13}$ rad/s; $\gamma_1=1.1\times10^{13}$ rad/s, $\gamma_2=2.2\times10^{13}$ rad/s, and
$\gamma_3=3.8\times10^{13}$ rad/s. The dashed lines 1, 2, and 3 show the reflectivities at the normal incidence of the respective
uncoated low-resistivity Si plates. As is seen in Fig.~8, the impact of graphene coating on the reflectivity quickly decreases
with increasing concentration of charge carriers and increasing frequency.

The dependences of the TM and TE reflectivities of the graphene-coated plate made of low-resistivity Si on the angle of incidence at low
frequencies can be  calculated by Eqs.~\eqref{eq:02},~\eqref{eq:20}, and~\eqref{eq:23}. The computational results at $\w=1$~meV$=1.5\times10^{12}$
rad/s, $T=300$K are shown in Fig.~9 by the lower and upper solid lines for the TM  and TE polarizations, respectively (the dashed
lines demonstrate the respective results for an uncoated low-resistivity
Si plate). In Figs. 9(a) and 9(b) the concentrations of charge carriers were chosen to be $N=5\times10^{14}$ and $5\times10^{17}\text{cm}^{-3}$,
respectively. As is seen in Fig.~9(a), for a lower conductivity uncoated Si plate the Brewster angle is equal to $\ti \approx 73.5^{\circ}$. For
the graphene-coated plate, however, the full polarization of the reflected light does not happen at any incidence angle (the minimum value
$\mathcal{R}_\tm \approx 0.022$ is achieved at $\ti\approx74.3^{\circ}$). For a higher conductivity Si [Fig.~9(b)] the full polarization
does not happen either for an uncoated or for a graphene-coated plate. The minimum values $\mathcal{R}_\tm \approx 0.15$ and 0.18 are achieved
at $\ti=84^{\circ}$ and $84.8^{\circ}$ for an uncoated and graphene-coated plates, respectively.

 At the end of this section, we note that for Si plates with the highest doping concentrations
 ($\sim 10^{20}\;\text{cm}^{-3}$)
there is no influence of graphene coating on the reflectivity properties. In the region of
high frequencies this fact was already discussed above, and at low frequencies the situation is the same as for metals
considered in Sec.~\ref{sec:ref_gm}.

\section{\label{sec:ref_cd}Conclusions and Discussion}

In the foregoing, we have developed theoretical description for the reflectivity properties of the graphene-coated plates made of different
materials (dielectric, metal and semiconductor). This theory is based on the Dirac model of graphene, which is described by the recently
found polarization tensor in (2+1)-dimensional space-time, allowing the analytic continuation to the real frequency axis~\cite{c27}.
The materials of the plate are described by the frequency-dependent complex index of refraction. In the framework
of this theory, the general formulas are obtained for the reflection coefficients and reflectivities of graphene-coated plates,
and their analytic asymptotic expressions at both high and low frequencies.

The developed theory is applied to the cases of graphene-coated dielectric, metal, and semiconductor plates. It is shown that for a dielectric
material (fused silica) in the region of optical and ultraviolet frequencies  the reflectivity of the graphene-coated plate at the normal
incidence is by several percent larger than of an uncoated one. In this frequency region, the graphene coating also results in some increase of the Brewster angle. In the asymptotic region of low frequencies, graphene coating  is shown to significantly (up to an order of magnitude)
increase the reflectivity of silica at the normal incidence. This effect is completely determined by the thermal corrections at room temperature.

For metals, the developed theory is illustrated by the examples of Au and Ni plates. In the case of high and moderate frequencies, we have shown that
the graphene coating decreases the plate reflectivity at the normal incidence. For low frequencies the graphene coating
does not influence the reflectivities of Au and Ni plates.

We have also applied the developed theory to graphene-coated semiconductor plates (Si with various concentrations of charge carriers).
For the high-resistivity Si plate, the influence of graphene coating at high frequencies achieves several percent similar to the case of a silica plate.
In the region of moderate and low frequencies, the graphene coating can significantly increase the reflectivity of Si plate at the normal incidence. For the low-resistivity Si the effect of graphene coating was investigated at different concentrations of charge carriers. It is shown
that at high frequencies the graphene coating does not influence the reflecting properties. In the region of low frequencies the impact of graphene
coating  on the reflectivity of Si plate at the normal incidence quickly decreases with increasing concentration of charge
carriers and increasing frequency. The dependences of the reflectivities on the angle of incidence were also computed.
According to our results, the graphene coating increases the reflectivities of low-resistivity Si  at low frequencies.
The full polarization of the reflected light is not achieved at any angle of incidence.

As was noted in Sec.~\ref{sec:intro}, the theory of reflectivity properties of graphene-coated material plates is of interest in many
technological applications using graphene coatings, such as the optical detectors, transparent conductors, anti-reflection surfaces in solar-cells etc.
In the future, it is planed to generalize the developed formalism to the case of graphene with nonzero
mass-gap parameter and chemical potential. It would be useful also to apply it to thin material plates (films) coated with graphene.

\section*{Acknowledgments}
The authors are grateful to V. M. Mostepanenko for several helpful discussions and useful suggestions.


\begin{figure}[t!]
\vspace*{-220pt}%
\includegraphics[width=.9\paperwidth]{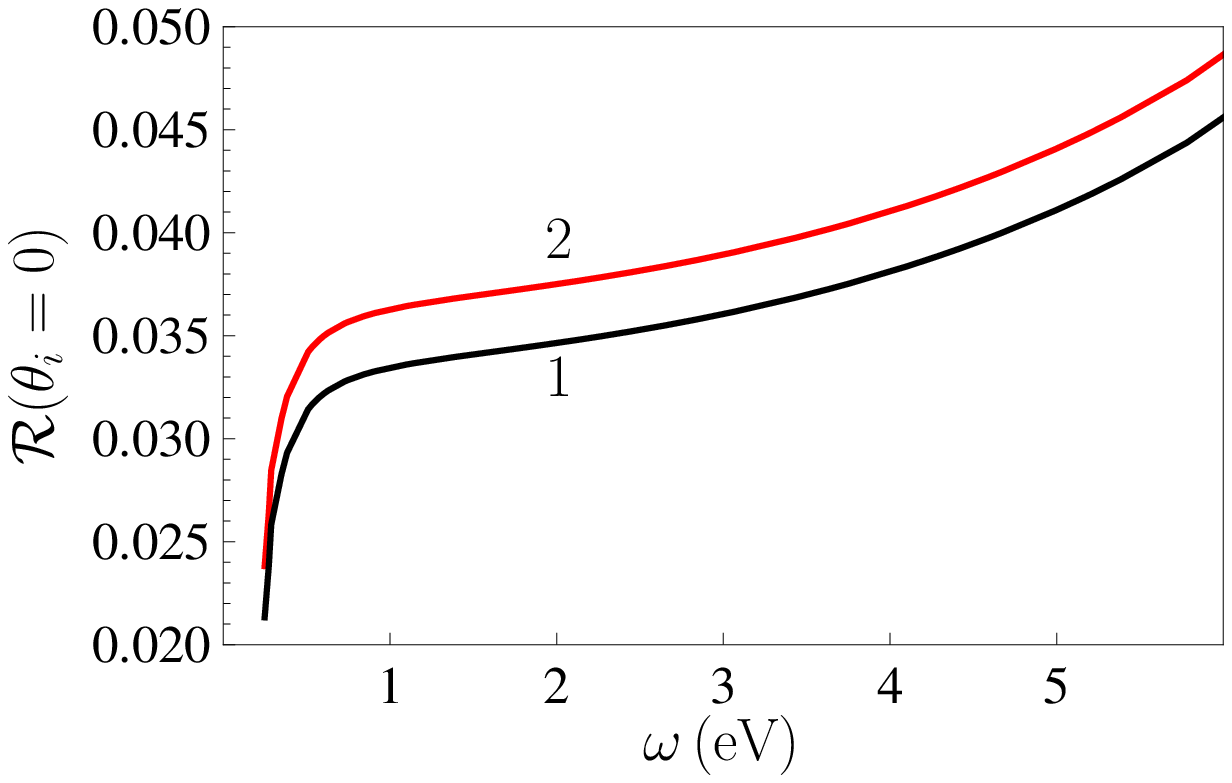}
\vspace*{-280pt}%
\caption{\label{fig:01}
The reflectivities of the graphene-coated and uncoated silica plates (the lines 2 and 1, respectively) at the normal incidence are shown
as functions of frequency at high frequencies.
}
\end{figure}

\begin{figure*}[tp]
\vspace*{-50pt}%
\includegraphics[width=.9\paperwidth]{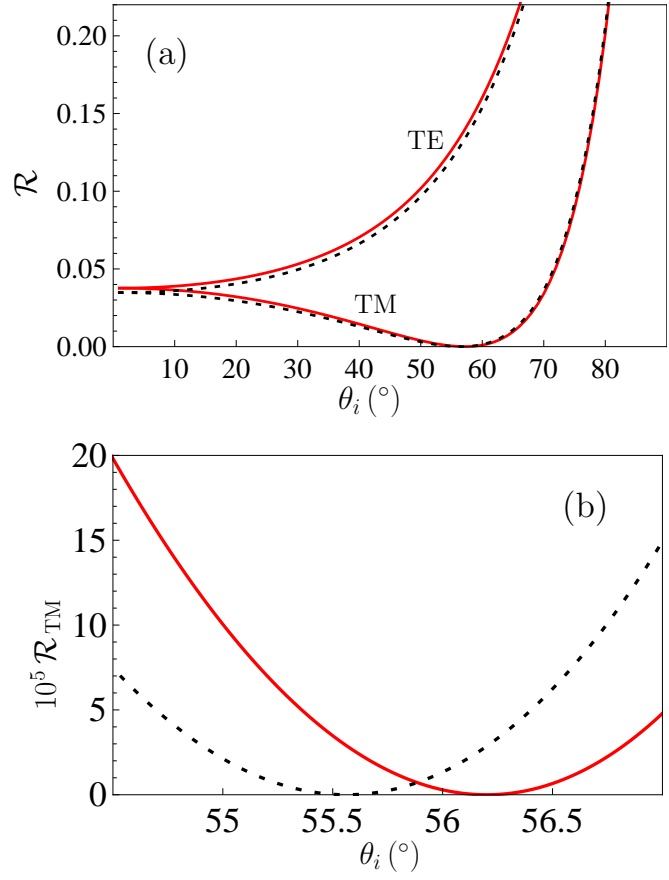}
\vspace*{-340pt}%
\caption{\label{fig:02} (a) The TM and TE reflectivities of the graphene-coated silica plate at $\w=3\times10^{15}$ rad/s are shown as functions of the
incidence angle by the lower and upper solid lines, respectively. The dashed lines show similar results for an uncoated silica plate.
(b) The TM reflectivities of graphene-coated and uncoated silica plates at $\w=3\times 10^{15}$ rad/s are shown by the solid and dashed lines, respectively,
in the vicinity of the Brewster angle.
}
\end{figure*}

\begin{figure*}[ht]
\vspace*{-220pt}%
\includegraphics[width=.9\paperwidth]{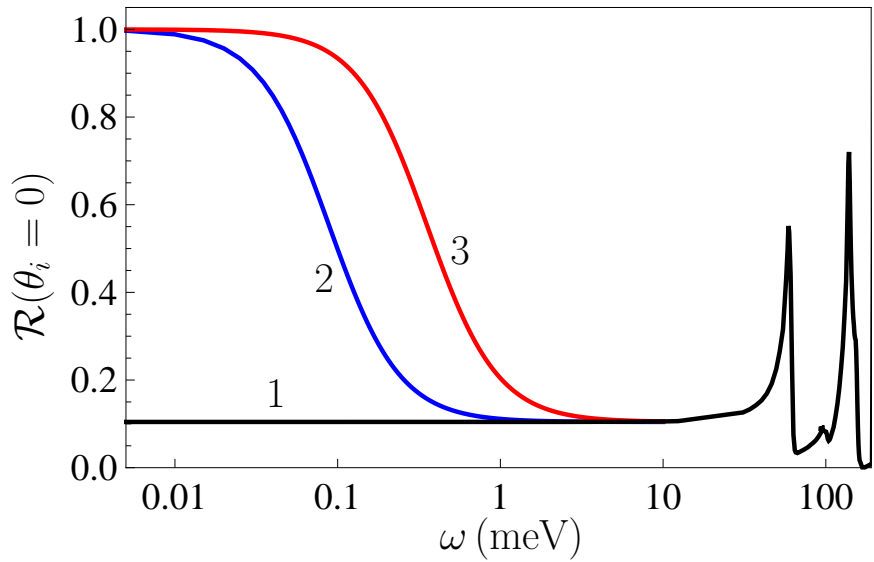}
\vspace*{-280pt}%
\caption{\label{fig:03} The reflectivities of the graphene-coated and uncoated silica plates (the lines 2, 3 and 1, respectively) at the normal
incidence are shown as functions of frequency at low frequencies. The lines 2 and 3 are plotted for the graphene-coated silica at $T=75$ K and
$T=300$ K, respectively.}
\end{figure*}

\begin{figure*}[ht]
\vspace*{-50pt}%
\includegraphics[width=.9\paperwidth]{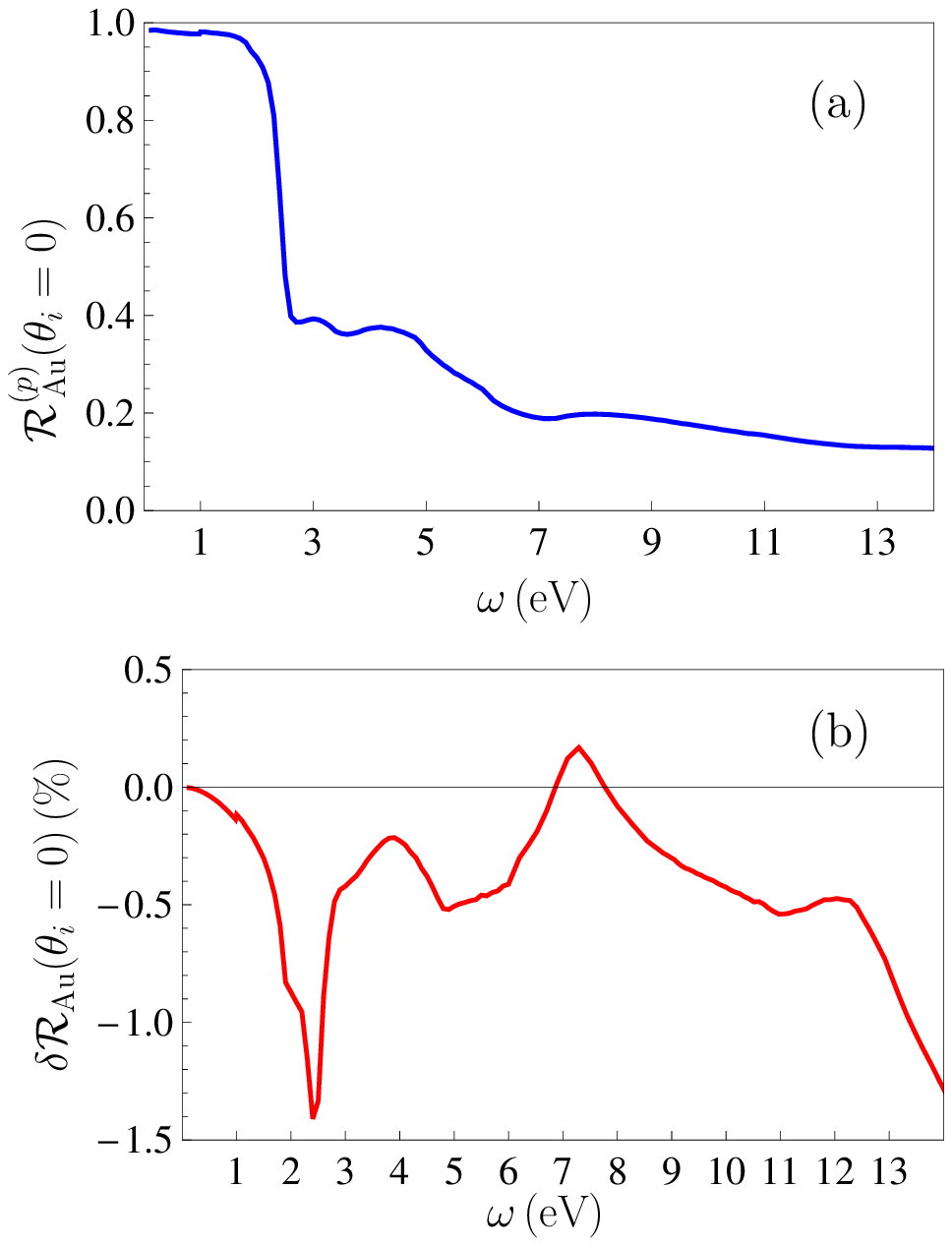}
\vspace*{-340pt}%
\caption{\label{fig:04} (a) The reflectivities of the uncoated Au plate and (b) the relative change of reflectivity of the graphene-coated Au plate at the normal
incidence are shown as functions of frequency.}
\end{figure*}

\begin{figure*}
\vspace*{-220pt}%
\includegraphics[width=.9\paperwidth]{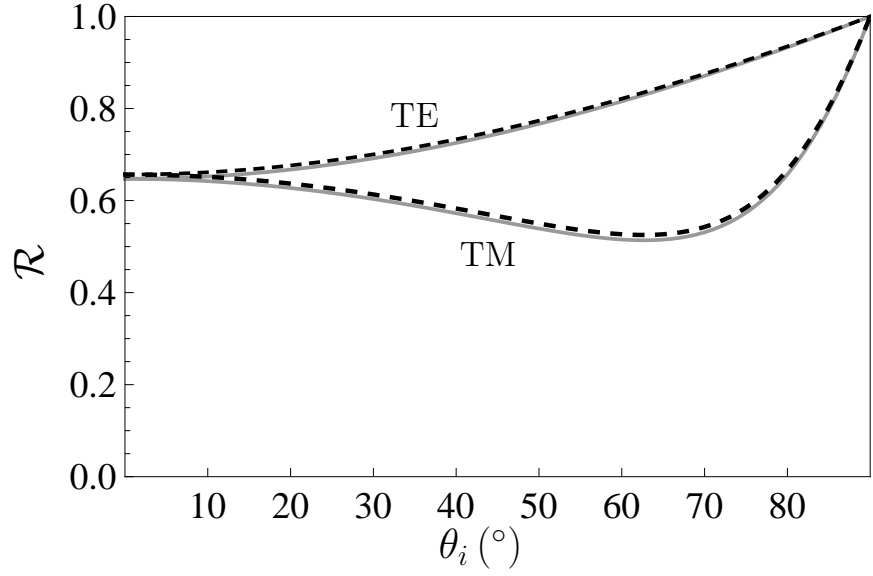}
\vspace*{-280pt}%
\caption{\label{fig:05} The TM and TE reflectivities of the graphene-coated Au plate at $\w=3.65\times10^{15}$ rad/s are shown as functions of
the incidence angle by the lower and upper solid lines, respectively. The dashed lines show similar results for an uncoated Au plate.}
\end{figure*}

\begin{figure*}
\vspace*{-50pt}%
\includegraphics[width=.9\paperwidth]{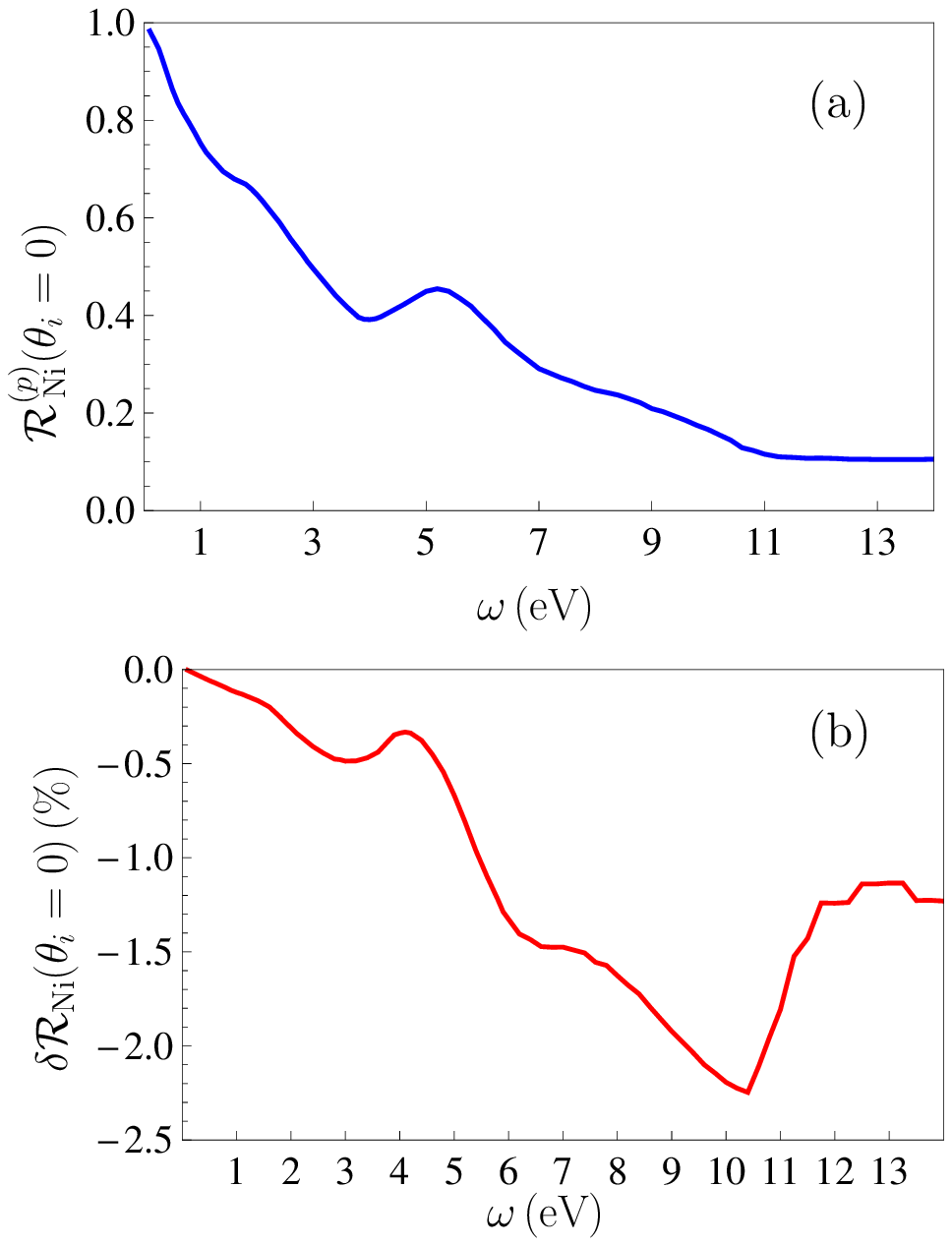}
\vspace*{-340pt}%
\caption{\label{fig:06} (a) The reflectivity of the uncoated Ni plate and (b) the relative change of reflectivity of the graphene-coated Ni plate at the normal incidence are shown as functions of frequency.}
\end{figure*}

\begin{figure*}
\vspace*{-80pt}%
\includegraphics[width=1.1\paperwidth]{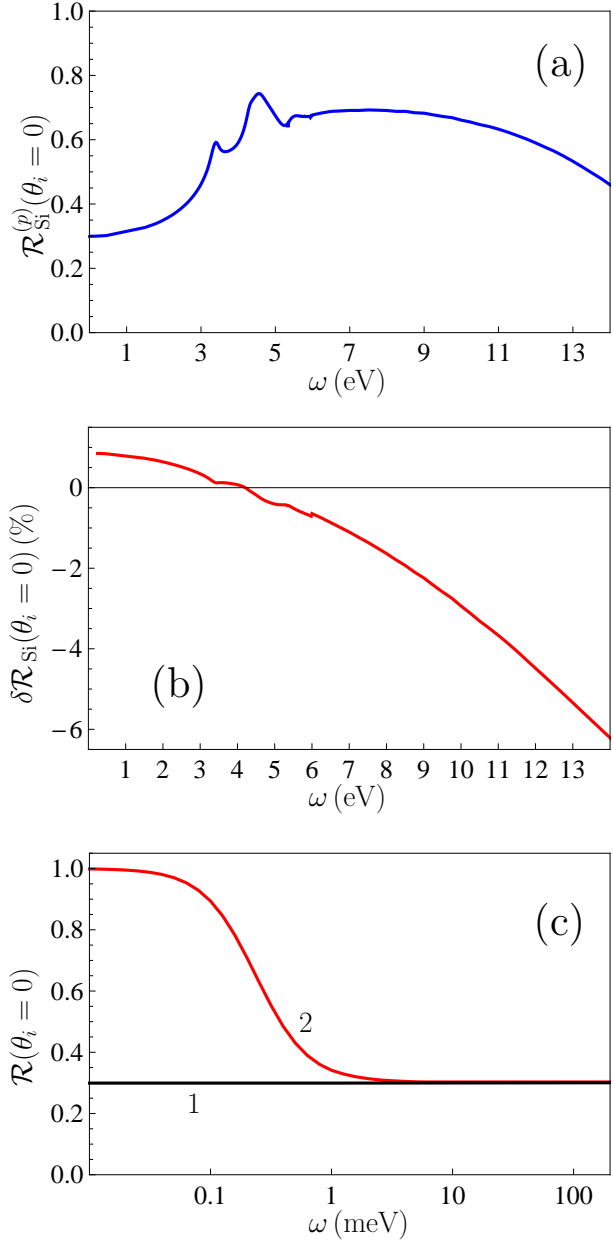}
\vspace*{-370pt}%
\caption{\label{fig:07} (a) The reflectivity of the uncoated high-resistivity Si plate, (b) the relative change of reflectivity of the high-resistivity
graphene-coated Si plate, and (c) the reflectivities of the high-resistivity graphene-coated Si plate at $T=300$ K and $T=0$ K
(the lines 2 and 1, respectively) at the normal incidence are shown as functions of frequency.}
\end{figure*}

\begin{figure*}
\vspace*{-220pt}%
\includegraphics[width=.9\paperwidth]{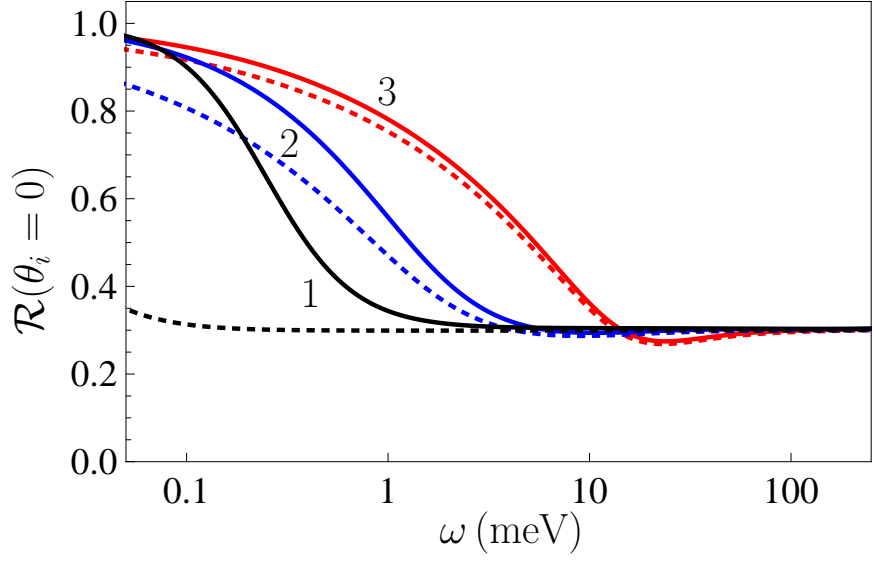}
\vspace*{-280pt}%
\caption{\label{fig:08} The reflectivities of the low-resistivity graphene-coated Si plates are shown at $T=300$ K by the solid
lines 1, 2, and 3 as functions of frequency for the concentrations of free charge carriers equal to $5\times10^{14}$, $5\times10^{16}$,
and $5\times10^{17} \text{cm}^{-3}$, respectively. The respective quantities for the uncoated plates are shown by the dashed lines 1, 2, and 3.}
\end{figure*}

\begin{figure*}
\vspace*{-60pt}%
\includegraphics[width=.9\paperwidth]{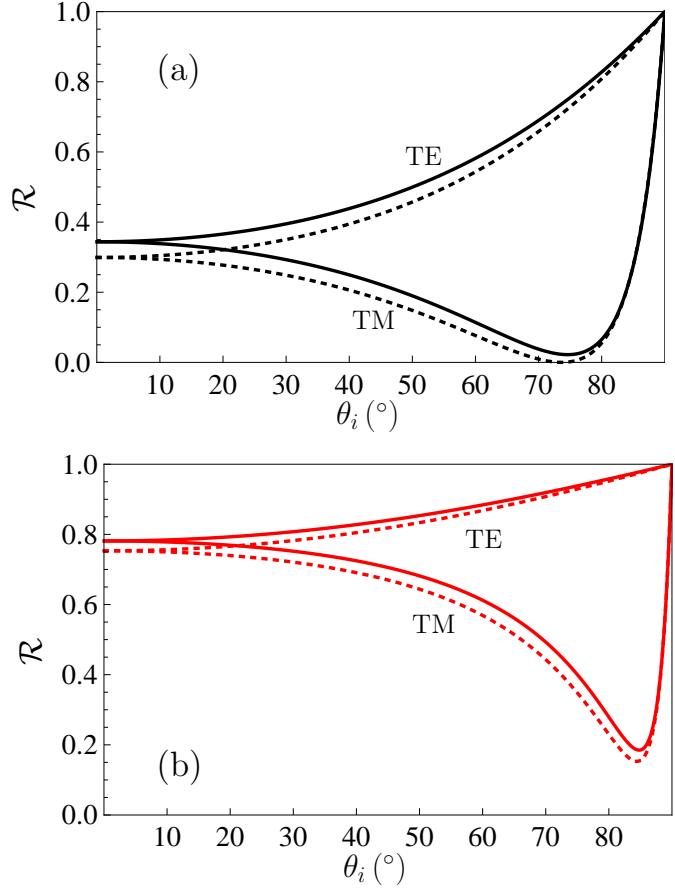}
\vspace*{-340pt}%
\caption{\label{fig:09}The TM and TE reflectivities of the graphene-coated low-resistivity Si plate at $\w=1.5\times10^{12}$ rad/s are shown at
$T=300$ K as functions of the incidence angle by the lower and upper solid lines, respectively, for the doping concentrations
(a) $N=5\times10^{14}\text{cm}^{-3}$ and
(b) $N=5\times10^{17} \text{cm}^{-3}$. The dashed lines show similar results for an uncoated Si.}
\end{figure*}
\end{document}